\documentstyle[11pt,epsf]{article}
\newcommand{\al}{\alpha}
\newcommand{\bt}{\beta}
\newcommand{\de}{\delta}

\newcommand{\Ga}{\Gamma}
\newcommand{\ga}{\gamma}
\newcommand{\be}{\begin{equation}}
\newcommand{\ee}{\end{equation}} 
\newcommand{\eei}{\end{equation}\indent\indent}
\newcommand{\bc}{\begin{center}}
\newcommand{\ec}{\end{center}}
\newcommand{\ber}{\begin{eqnarray}}
\newcommand{\ear}{\end{eqnarray}}
\newcommand{\ba}{\begin{array}}
\newcommand{\ea}{\end{array}}
\newcommand{\fr}{\frac}
\newcommand{\ta}{\tau}
\newcommand{\vb}{\verb}

\newcommand{\na}{\nabla}

\newcommand{\p}{\partial}

\def\case#1/#2{\textstyle\frac{#1}{#2} }
\begin{document}
\title{An Expansion Term In Hamilton's Equations.}
\author{Mark D. Roberts, \\\\
Department of Mathematics and Applied Mathematics, \\ 
University of Cape Town, South Africa\\\\
roberts@gmunu.mth.uct.ac.za}
\date{\today}
\maketitle
\bc Published:  Europhysics Letters {\bf 45}(1991)26-31 \ec
\bc             http://www.edpsciences.com/docinfos/EURO/OnlineEURO.html  \ec
\bc Epreprint:  http://xxx.lanl.gov/abs/gr-qc/9810090   \ec
\bc Comments:   Final Version,  no diagrams,  LaTex2e,
    small changes from previous eprint and from the published version.  \ec
\bc 1999 PACS Physics and Astronomy Classification Scheme: \ec
\bc http://publish.aps.org/eprint/gateway/pacslist  \ec
   \bc   11.10.Ef,  04.60+Ds,  52.60+h,  03.70+k  \ec
\bc 1991 Mathematical Reviews Classification Scheme:  \ec
\bc http://www.ams.org/msc  \ec
   \bc   81S99,  83C45,  83C43 \ec
\bc Keywords: \ec 
\bc Hamilton's Equations,  Preferred Vector Field, \ec
\bc Expansion, Sympletic Geometry.\ec
\newpage
\begin{abstract}
For any given spacetime the choice of time coordinate is undetermined.   
A particular choice is the absolute time associated with a preferred 
vector field.   Using the absolute time Hamilton's equations are
\newline
\bc
$-\frac{\delta H_{c}}{\delta q}=\dot{\pi}+\Theta\pi,$
\ec
\bc
$+\frac{\delta H_{c}}{\delta \pi}=\dot{q}$,
\ec
where $\Theta = V^{a}_{.;a}$ 
is the expansion of the vector field.   Thus 
there is a hitherto unnoticed term in the expansion of the preferred vector 
field.     Hamilton's equations can be used to describe fluid motion.  In this 
case the absolute time is the time associated with the fluid's co-moving 
vector.  As measured by this absolute time the expansion term is present.   
Similarly in cosmology,  each observer has a co-moving vector and 
Hamilton's equations again have an expansion term.    It is necessary to 
include the expansion term  to quantize systems such as the above by the 
canonical method of replacing Dirac brackets by commutators.
Hamilton's equations in this form do not have a corresponding sympletic
form.   Replacing the expansion by a particle number 
$N\equiv exp(-\int\Theta d \ta)$
and introducing the particle numbers conjugate momentum $\pi^{N}$ the standard 
sympletic form can be recovered with two extra fields $N$ and $\pi^{N}$.
Briefly the possibility of a non-standard sympletic form and 
the further possibility
of there being a non-zero Finsler curvature corresponding to this are 
looked at.
\end{abstract}
\newpage
\section{Introduction} \label{sec:intro}   
One of the problems with the standard treatment of Hamilton's equations
is that it assumes that a preferred time can be "taken out".  
This is not a restriction in special relativity 
or in static general relativity;  
however in a non-static spacetime one wants to define time 
with respect to a given preferred vector field.   
Below it is shown that defining time by this specific preferred vector field,  
the expansion of the vector field has to 
be included in Hamilton's equations.   
The expansion term also occurs in the Poisson bracket 
and the equation for ordinary time evolution of an arbitary variable.   
The type of systems where the expansion term is important 
are perfect fluids, where the preferred vector field is that of a co-moving
element of the fluid;  and in cosmology where the preferred 
vector field is that of an observer co-moving with the expansion.  
If one wants to quantize such systems one has to 
replace the Poisson (or for constrained systems the Dirac) bracket
by a quantum commutator.   After this is done the term in the expansion
remains.   For a perfect fluid quantization has been done with 
expansion present Roberts (1998) \cite{bi:mdr2}.   
Quantization of cosmology,  Tipler (1986)
\cite{bi:tipler} is usually approached via the ADM 
(Arnowitt,  Deser,  and Misner (1962)) formalism \cite{bi:ADM}.
In this time is taken out so that the normal to the surface has no expansion,
vorticity,  or acceleration.   
It is possible that this can be generalized using the parametric 
manifold approach of Boersma and Dray (1995) \cite{bi:BD}.
The inclusion of a normal vector with 
expansion would require use of Hamilton's equations with the expansion 
term presented here.   Expansion is the main qualitative property in
cosmology so that the addition of ths term is of importance. 
The additional expansion term was first found for 
the Hamiltonian description of particular fluids and then generalized. 
The expansion term might be directly derivable by different methods
using Peierls (1952) \cite{bi:peierls} brackets.
The manner in which time is usually taken out is simply to label one coordinate
as $t$.   If there is a preferred vector field $V^{a}$ present it is associated
with a preferred time via the absolute derivative
\be
\dot{X}_{abc\ldots}=
\fr{D}{d \ta}X_{abc\ldots}=V^{e}\na_{e}X_{abc\ldots}=V^{e}X_{abc\ldots;e}.
\label{eq:absdev}
\ee
The usual approach is recovered by letting $V^{a}=(1,0,0,0)$.
In general $V^{a}$ has a more complex structure than this and can be decomposed
into the shear,  vorticity,etc\ldots,  see for example
Hawking and Ellis (1973) \cite{bi:HE} p.83.
Additions to Hamilton's equations have also been used to overcome
conceptual problems in quantum mechanics
Ghirardi,  Rimini,  and Weber (1986) \cite{bi:GRW}. 
\section{Hamilton's equations}
The canonical Hamiltonian $H_{c}$ for fields is
\be
H_{c}=\int\sqrt{-g}~d^{4}x~H_{d}
     =\int\sqrt{-g}~d^{4}x(\pi^{i}\dot{q}_{i}-L),
\label{eq:canham}
\ee
where $H_{d}$ is the Hamiltonian density.
The corresponding canonical phase space Lagrange action is
\be
I_{ph}=\int\sqrt{-g}~d^{4}x(\pi^{i}\dot{q}_{i}-H_{d}).
\ee
Varing $I_{ph}$
\ber
\delta I_{ph}&=&\int\sqrt{-g}~d^{4}x[
              \pi^{i} \de (\fr{d \dot{q}_{i}}{d \ta})
            +\fr{d \dot{q}_{i}}{d \ta}\de \pi^{i} 
            -(\fr{\p H_{d}}{\p q^{i}})\de q_{i}\nonumber\\
            &-&(\fr{\p H_{d}}{\p q^{i}_{a}})\de q^{i}_{a}
            -(\fr{\p H_{d}}{\p \pi^{i}})\de \pi^{i}
            -(\fr{\p H_{d}}{\p \pi^{i}_{a}})\de \pi^{i}_{a}],
\ear
replacing $\frac{D}{d\tau}$ by $V^{a}\nabla_{a}$,  as in equation
\ref{eq:absdev}  
taking $\delta V^{a}=0$,  interchanging infinitesimals thus
\ber
\pi_{i}\left(\frac{\delta d\dot{q}^{i}}{d\tau}\right)  
&=&\pi_{i}\delta(V^{a}q^{i}_{.;a})
=\pi_{i}V^{a}\delta (q^{i}_{.;a})\nonumber\\ 
&=&\pi V^{a} \nabla_{a}(\delta q^{i})
=\nabla_{a}(V^{a}\pi_{i}\delta q^{i})-(\delta q^{i})\nabla_{a}(\pi_{i}V^{a}),
\ear
and disgarding the surface term $\na_{a}(V^{a}\pi_{i}\de q^{i})$ leaves
\ber
\delta I_{ph}=\int\sqrt{-g}d^{4}x[
&\delta q^{i}&\left(-\nabla_{a}(\pi_{i}V^{a})
-\frac{\p H_{d}}{\p q^{i}}+\nabla_{a}
\frac{\p H_{d}}{\p q^{i}_{.;a}}\right)\nonumber\\
+&\delta \pi^{i}&\left(V^{a}\nabla_{a}q_{i}
-\frac{\p H_{d}}{\p \pi^{i}}+\nabla_{a}
\frac{\p H_{d}}{\p \pi^{i}_{.;a}}\right)].
\ear
For the $\delta q^{i}$ term to vanish
\be
-\frac{\delta H_{c}}{\delta q^{i}}
=-\frac{\p H_{d}}{\p q^{i}}+\nabla ( \frac{\p H_{d}}{\p q^{i}_{.;a}})
=\nabla_{a}(\pi^{i}V^{a})
=\dot{\pi}^{i}+\Theta \pi^{i}
\label{eq:heq}
\ee
For the $\delta \pi^{i}$ term to vanish
\be
+\frac{\delta H_{c}}{\delta \pi^{i}}
=+\frac{\p H_{d}}{\p \pi^{i}}-\nabla ( \frac{\p H_{d}}{\p \pi^{i}_{.;a}})
=V^{a}\nabla_{a}q^{i}
=\dot{q}^{i}
\label{eq:hep}
\ee
Equations \ref{eq:heq} and \ref{eq:hep} generalize Hamilton's equations.
The equations are uneffected if $\pi^{i}$ and $q^{i}$ have spacetime 
indices $\pi^{i}\rightarrow \pi^{i}_{ab\ldots},
q^{i}\rightarrow q^{i}_{ab\ldots}$ and so on.   
The standard Poisson bracket is defined as
\be
\vb+{+X,Y\vb+}+\equiv
\fr{\de X}{\de q_{i}}\fr{\de Y}{\de \pi^{i}}
-\fr{\de X}{\de \pi_{i}}\fr{\de Y}{\de q^{i}}.
\ee
Equations \ref{eq:heq} and \ref{eq:hep} suggest 
alternative Poisson bracket to the standard bracket,
\ber
\vb+{+A,B\vb+}+_{A}
&=&\left(\Theta\pi_{i}+\fr{\de A}{\de q_{i}}\right)
\fr{\de B}{\de \pi^{i}}
-\fr{\de A}{\de \pi^{i}}
\left(\Theta \pi_{i}+\fr{\de B}{\de q_{i}}\right)\nonumber\\
&=&\vb+{+A,B\vb+}+ 
+\Theta\pi_{i}\left(\fr{\de B}{\de \pi^{i}}
                    -\fr{\de A}{\de \pi^{i}}\right),
\ear
containing the expansion $\Theta$.  The alternative bracket 
does not give $\vb+{+\pi_{i},q^{i}\vb+}+=1$, also for particular 
constrained systems the correct Lagrange multipliers are not recovered. 
Thus the standard bracket is preferred.  
Perhaps Peierls' brackets \cite{bi:peierls} can show that the 
alternative bracket was not suitable by a general method.   
The time evolution of an arbitrary 
variable $X(\ta,q^{i},\pi^{i})$ is given by
\be
\fr{d X}{d \ta} 
=\fr{\p X}{\p \ta}+\dot{q}^{i}\fr{\de X}{\de q^{i}}
                     +\dot{\pi}^{i}\fr{\de X}{\de \pi^{i}},
\ee
substituting $q^{i}$ and $\pi^{i}$ using equations
\ref{eq:heq} and \ref{eq:hep} gives the time evolution
\ber
\fr{dX}{d\ta}
&=&\fr{\p X}{\p \ta}+\vb+{+X,H_{c}\vb+}+
-\Theta \pi^{i}\fr{\de X}{\de \pi^{i}}\nonumber\\
&=&\fr{\p X}{\p \tau}+\vb+{+X,H_{c}\vb+}+_{A}
-\Theta \pi^{i}\fr{\de H_{c}}{\de \pi^{i}}.
\ear
Note in particular that the time evolution of the canonical Hamiltonian $H_{c}$
is no longer soley given by its partial derivative with respect to time and is
\be
\fr{d H_{c}}{d \ta}=\fr{\p H_{c}}{\p \ta}
-\Theta\pi^{i}\fr{\de H_{c}}{\de \pi^{i}}.
\label{eq:evolh}
\ee
\section{Sympletic Structure}
Following the standard,  see for example Wald (1994), \cite{bi:wald} p.11
procedure for producing the sympletic form let
\be
y^{\mu}=(q^{1},\dots,q^{n};\pi^{1},\dots,\pi^{n})
\label{eq:yyy}
\ee
to produce the equations
\be
\fr{dy^{\mu}}{d\tau}=\sum_{\nu=1}^{2n}\Omega^{\mu\nu}
\fr{\de H_{c}}{\de y^{\nu}},
\label{eq:symat}
\ee
when $\nu=\mu+n$
\ber
\Omega^{\mu\nu}&=&1,\nonumber\\
\Omega^{\nu\mu}&=&-\fr{\dot{\pi}^{\mu}}{\dot{\pi}^{\mu}+\Theta\pi^{\mu}},
\ear
otherwise $\Omega=0$.   This is not a sympletic matrix as (unless
$\Theta=0$) it is not antisymmetric.   Another problem is that the 
expansion $\Theta$ is not explicitly defined in term of the fields
$q^{i}$ and the momenta $\pi^{i}$.   Also recall that for sympletic spaces 
the Hamiltonian is usually conserved,  Arnol\`{d} 1978 \cite{bi:arnold} p.207,
but from equation \ref{eq:evolh}, $\fr{dH_{c}}{d \ta}$ is not necessarily 
zero here.

In order to produce a sympletic structure first remove the expansion $\Theta$
so that there are only fields and momenta present.
By anology with fluids,  Roberts(1997) \cite{bi:mdr1}, 
let $\Theta=-\frac{\dot{N}}{N}$ 
where $N$ is a particle number.   The existing $2i$ equations are
\ber
-\fr{\de H_{c}}{\de q^{i}}
&=&\dot{\pi}^{i}+\fr{\dot{N}}{N}\pi^{i}=
N\dot{\pi'}^{i}\;,~~~~~~~~~~~
\dot{\pi'}^{i}\equiv\fr{\pi^{i}}{N},\nonumber\\
+\fr{\de H_{c}}{\de \pi^{i}}&=&\dot{q}^{i}.
\label{eq:exeq}
\ear
N is a new field so it also has to be added to the Hamiltonian density
to give the new Hamiltonian density
\be
H_{a}=H_{d}+\pi_{N}\dot{N}=\pi^{i+1}\dot{q}_{i+1}-L,
\label{eq:hamn}
\ee
and new canonical Hamiltonian $H_{A}$.
In addition to the 2i equations described by \ref{eq:exeq} 
there are two new equations
\ber
-\fr{\de H_{A}}{\de N}&=&\dot{\pi}^{N}-\frac{\dot{N}}{N}\pi^{N}
=N\dot{\pi}'^{N}\;,~~~~~~~~~~\pi'^{N}\equiv\fr{\pi^{N}}{N}\nonumber\\
+\fr{\de H_{A}}{\de \pi^{N}}&=&\dot{N},
\label{eq:heqn}
\ear
From equations \ref{eq:hamn} and \ref{eq:heqn} we can now define
\be
q'^{i}\equiv\int d\ta\left(\fr{\dot{q}^{i}}{N}\right),
\ee
and
\be
z^{\mu}\equiv(q'^{1},\ldots,q'^{n},N;\pi'^{1},\ldots,\pi'^{n},\pi'^{N}),
\ee to produce equations of the same form as \ref{eq:symat} with $z^{\mu}$
replacing $y^{\mu}$ and
\be
\Omega^{\mu \nu}=N\;,~~~~~~~~~~~~\Omega^{\nu \mu} = -N,
\ee
otherwise $\Omega=0$.  This sympletic matrix describes the system and is two
dimensions higher than the original dimension,  also it is $N$ times the 
standard form.   To produce a sympletic matrix of the standard form 
rather than $N$ times it,  leave $q^{i}$ in its original form and define
\be
\pi''^{i}\equiv\int d\ta\left(\dot{\pi}^{i}+\fr{\dot{N}}{N}\pi^{i}\right).
\ee

A possibility is that the extra structure present because of the expansion
can lead to a non standard form of the sympletic matrix and that this in 
turn would give non-vanishing Finsler curvature,  Rund (1959) \cite{bi:rund},
of some geometric object.   In order to attempt investigation of this
{\bf first} try adding a term $\ga(q,\dot{q},\pi,\dot{\pi})$ 
to the Hamiltonian density $H_{a}$ to give a new Hamiltonian density
\be
H_{\gamma}=\pi^{i+1}\dot{q}_{i+1}-L+\gamma(q,\dot{q},\pi,\dot{\pi}),
\ee
and a new canonical Hamiltonian $H_{\Ga}$.
Hamilton's equations become
\ber
-\fr{\de H_{\Ga}}{\de q^{i}}
&=&\dot{\pi}^{i}-\fr{\dot{N}}{N}\pi^{i}+\fr{\p \ga}{\p q^{i}}
-\left(\fr{\p \ga}{\p \dot{q}}\right)^{\cdot}
+\fr{\dot{N}}{N}\fr{\p \ga}{\p \dot{q}^{i}},\nonumber\\
+\fr{\de H_{\Ga}}{\de \pi^{i}}
&=&\dot{q}^{i} +\fr{\p\ga}{\p \pi^{i}}
-\left(\fr{\p \ga}{\p \dot{\pi}^{i}}\right)^{\cdot}
+\fr{\dot{N}}{N}\fr{\p \ga}{\p \dot{\pi}^{i}}.
\ear
Taking an ansatz for $\ga$ of the form
\ber
\ga=k_{1}+k_{2}q&+&k_{3}\pi+k_{4}q\pi\nonumber\\
   +k_{5}\dot{q}+k_{6}q\dot{q}&+&k_{7}\pi\dot{q}+k_{8}q\pi\dot{q}\nonumber\\
   +k_{9}\dot{\pi}+k_{10}q\dot{\pi}&+&k_{11}\pi\dot{\pi}+k_{12}q\pi\dot{\pi},
\ear
it is immediately apparent that $k_{1}=k_{2}=k_{3}=k_{4}=k_{5}=k_{7}=k_{9}=0$
and can be shown that nothing of this form will produce an antisymmetic 
sympletic matrix.   {\bf Secondly} recall,   Hargreaves (1908) 
\cite{bi:hargreaves} that in fluid mechanics 
variation of the Lagrangian and Hamiltonian is achieved via the infinitesimal 
first law of thermodynamics,  typically
\be
\delta L=\delta p = N\delta h -NT\delta s + \sum X_{\mu}\de x^{\mu},
\label{eq:1stlaw}
\ee
where p is the pressure,  N the particle number,  h the enthalphy,  
T the temperature, and s the entropy.   $X_{\mu}$ are any additional
intensive variables,  $x^{\mu}$ are any additional extensive variables.
For definintions of intensive and extensive variables 
see Adkins (1983) \cite{bi:adkins};
briefly if an intensive varaible is doubled the mass remains the same,
but if an extensive variable is doubled the mass doubles.
In particular term $\mu \de N$ can be added,  
where $\mu$ is the chemical potential.   
By analogy with the first conservation law \ref{eq:1stlaw} choose
new Hamilitonian density $H_{e}$ which can be expressed infinitesimally 
in the form
\be
\de H_{e}=\de H_{n}-\al\times(\de \bt).
\ee
It immediately follows that the choice 
$\al=\fr{\dot{\pi}^{N}}{\pi^{N}}q^{i},\bt=\pi_{i}$ gives the equations
\ber
-\fr{\de H_{E}}{\de q^{i}}
&=&\dot{\pi}^{i}-\fr{\dot{N}}{N}\pi^{i},\nonumber\\
+\fr{\de H_{E}}{\de \pi^{i}}
&=&\dot{q}^{i}-\fr{\dot{\pi}^{N}}{\pi^{N}}q^{i},
\ear
where i ranges form $1$ to $n+1$ so as to include the particle number $N$ and 
its associated momentum $\pi^{N}$.   
Using equations \ref{eq:yyy} and \ref{eq:symat}
with $n$ replaced with $n+1$ to try and remove the time derivatives and produce
a sympletic matrix one instead arrives at
\ber
\fr{d q^{i}}{d \ta}&=&\fr{\de H_{E}}{\de \pi_{i}}+xq^{i}\left(
-\fr{1}{\pi^{N}}\fr{\de H_{E}}{\de N}+\fr{1}{N}\fr{\de H_{E}}{\de \pi^{N}}
\right),\nonumber\\
\fr{d \pi^{i}}{d \ta}&=&-\fr{\de H_{E}}{\de q_{i}}-y\pi^{i}\left(
-\fr{1}{N}\fr{\de H_{E}}{\de \pi^{N}}+\fr{1}{\pi^{N}}\fr{\de H_{E}}{\de N},
\right)
\ear
where $x,y$ are the number of iterations used to remove $\dot{N}$
and $\dot{\pi}^{N}$.   Taking $x$ and $y$ to be finite,   
\be
y^{\mu}=(q^{i},\pi^{i},N,\pi^{N}),
\ee
and equation \ref{eq:symat} gives in place of $\Omega^{\mu \nu}$ the matrix
\ber
M^{\mu\nu}=
\left(\begin{array}{c}
0~~~~~~~~1~~~~~~~~-\fr{xq^{i}}{\pi^{N}}~~~~~~~\fr{xq^{i}}{N}~\\
-1~~~~~~~~0~~~~~~~~~-\fr{y\pi^{i}}{\pi^{N}}~~~~~~~\fr{y\pi^{i}}{N}~\\
0~~~~~~~0~~~~~~~~-x\fr{N}{\pi^{N}}~~~~~~~~1+x~\\
0~~~~~~~0~~~~~~-1-x~~~~~~~~~y\fr{\pi^{N}}{N}
\end{array}\right),
\ear
which has trace $y\fr{\pi^{N}}{N}-x\fr{N}{\pi^{N}}$
and determinant $1+x+y$,
however it is not anti-symmetric and so is not sympletic.
\section{Conclusion}
It was shown that when ``time'' used is that of a preferred vector field
that Hamilon's equations gain a term in the expansion of the vector field.
The implications of this for sympletic structure were discussed.
\section{Acknowledgements}
This work has been supported in part by the South African Foundation
for Research and Development (FRD).

\end{document}